\documentclass[journal,twoside]{IEEEtran}
\usepackage{amsmath,amssymb,amsfonts}
\usepackage{algorithmic}
\usepackage{graphicx}
\usepackage{textcomp}
\usepackage{siunitx}
\usepackage{booktabs}
\usepackage[colorlinks=true, linkcolor=blue, citecolor=blue, urlcolor=blue]{hyperref}
\usepackage[nolist]{acronym}
\usepackage[style=ieee,maxnames=6,minnames=1,doi=true,url=false,isbn=false,date=year]{biblatex}

\begin{document}
\title{Beam Hardening Correction in Clinical X-ray Dark-Field Chest Radiography using Deep Learning-Based Bone Segmentation}
\author{Lennard Kaster$^{*}$, Maximilian E. Lochschmidt$^{*}$, Anne M. Bauer, Tina Dorosti, Sofia Demianova, Thomas Koehler, Daniela Pfeiffer, and Franz Pfeiffer
\thanks{We acknowledge financial support through the European Research Council (ERC Synergy Grant SmartX, SyG 101167328), and the Free State of Bavaria under the Excellence Strategy of the Federal Government and the States, as well as by the Technical University of Munich – Institute for Advanced Study.}
\thanks{This work involved human subjects or animals in its research. Approval of all ethical and experimental procedures and protocols was granted by the Ethics Commission of the Medical Faculty, Technical University of Munich, Germany, under Reference No. 166/20S, and the National Radiation Protection Agency (Bundesamt für Strahlenschutz) under Application No. Z5-224.}
\thanks{Lennard Kaster and Maximilian E. Lochschmidt contributed equally to this work, corresponding author: Lennard Kaster (email: lennard.kaster@tum.de).}
\thanks{Lennard Kaster, Maximilian E. Lochschmidt, Tina Dorosti, Sofia Demianova, and Franz Pfeiffer are with the Chair of Biomedical Physics, Department of Physics, School of Natural Sciences, Technical University of Munich, Garching, Germany and the Munich Institute for Biomedical Engineering, Technical University of Munich, Garching, Germany}
\thanks{Lennard Kaster, Maximilian E. Lochschmidt, Tina Dorosti, Sofia Demianova, Daniela Pfeiffer, and Franz Pfeiffer are with the Institute for Diagnostic and Interventional Radiology, School of Medicine and Health, TUM Klinikum, Technical University of Munich (TUM), Munich, Germany.}
\thanks{Franz Pfeiffer, Thomas Koehler, and Daniela Pfeiffer are with the Institute for Advanced Study, Technical University of Munich, Garching, Germany.}
\thanks{Anne Bauer was with the Chair of Biomedical Physics, Department of Physics, School of Natural Sciences, Technical University of Munich, Garching, Germany and the Munich Institute for Biomedical Engineering, Technical University of Munich, Garching, Germany and with the Department of Diagnostic and Interventional Radiology, TUM University Hospital, Klinikum rechts der Isar, Technical University of Munich, Munich, Germany.}
\thanks{Thomas Koehler is with Philips Innovative Technologies, Hamburg, Germany.}}
\maketitle
\begin{abstract}
Dark-field radiography is a novel X-ray imaging modality that enables complementary diagnostic information by visualizing the microstructural properties of lung tissue. 
Implemented via a Talbot-Lau interferometer integrated into a conventional X-ray system, it allows simultaneous acquisition of perfectly temporally and spatially registered attenuation-based conventional and dark-field radiographs.
Recent clinical studies have demonstrated that dark-field radiography outperforms conventional radiography in diagnosing and staging pulmonary diseases. 
However, the polychromatic nature of medical X-ray sources leads to beam-hardening, which introduces structured artifacts in the dark-field radiographs, particularly from osseous structures. 
 This so-called beam-hardening-induced dark-field signal is an artificial dark-field signal and causes undesired cross-talk between attenuation and dark-field channels.
 This work presents a segmentation-based beam-hardening correction method using deep learning to segment ribs and clavicles. Attenuation contribution masks derived from dual-layer detector computed tomography data, decomposed into aluminum and water, were used to refine the material distribution estimation. 
The method was evaluated both qualitatively and quantitatively on clinical data from healthy subjects and patients with chronic obstructive pulmonary disease and COVID-19. The proposed approach reduces bone-induced artifacts and improves the homogeneity of the lung dark-field signal, supporting more reliable visual and quantitative assessment in clinical dark-field chest radiography.
\end{abstract}


\section{Introduction}
\label{sec:introduction}
\IEEEPARstart{D}{ark-field} radiography is a pioneering imaging technique that enables visualization of micro-structural properties of the sample under investigation that are otherwise inaccessible \cite{Yashiro2010}, such as the lung's alveolar structure \cite{Bech2010, Schleede2012}. Compared to conventional radiography, clinical studies have shown the superiority of dark-field radiography for diagnosing and staging pulmonary diseases, such as COVID-19 and chronic obstructive pulmonary disease (COPD) \cite{Willer2021, Urban2022, Urban2024, Frank2022, Gassert2025}. 
The dark-field signal is generated from ultra-small-angle scattering at material interfaces, providing complementary information to the attenuation-based conventional radiograph. Healthy pulmonary tissue generates a strong dark-field signal due to its numerous alveoli, representing scattering air-tissue interfaces \cite{Gassert2021, Willer2021}. 
Pathological changes such as alveolar destruction, inflammation, or masses reduce these interfaces, resulting in a diminished dark-field signal \cite{Willer2021, Schleede2012, Hellbach2015, Scherer2017, Hellbach2018, Urban2022, Gassert2025}.
In contrast, non-microstructured tissues such as bone and soft tissue contribute minimally to the dark-field signal, making the modality highly specific to pulmonary microstructure.
 \\
To detect the ultra-small-angle scattering with clinical X-ray sources, as typically employed in radiography and CT systems, a Talbot-Lau interferometer is positioned within the beam path \cite{Pfeiffer2006,Pfeiffer2008}. 
The Talbot-Lau interferometer, comprising three gratings denoted as $\text{G}_0$, $\text{G}_1$, and $\text{G}_2$, allows for the simultaneous measurement of attenuation, dark-field, and differential phase-contrast signals. 
The source grating $\text{G}_0$ enables the utilization of a conventional source with a spot size significantly larger than the grating periods. The phase grating $\text{G}_1$ serves as the irradiated periodic pattern, generating an intensity pattern on the detector \cite{Momose2003}. However, the resolution of clinical flat-panel detectors is insufficient to directly resolve this fine intensity pattern.
Therefore, an analyzer grating, G2, with a period matching the intensity pattern, is positioned directly in front of the detector to overcome this limitation.
The mean intensity decreases when the intensity pattern is attenuated, as observed in conventional radiographs and CT scans. The dark-field signal diminishes the visibility of the intensity pattern, affecting the relative amplitude or contrast. 
The phase shift of the pattern represents the phase contrast signal.
By capturing different relative grating positions for each pixel, the attenuation, dark-field, and (differential) phase-contrast signals can be simultaneously extracted for each pixel \cite{Pfeiffer2008}. This capability enhances the potential of the Talbot-Lau interferometer in clinical X-ray dark-field chest radiography due to the perfectly spatially and temporally registered complementary information. \\
It is well-known from conventional X-ray radiographs and CT scans that strongly attenuating samples can cause artifacts impairing image quality and diagnostic accuracy \cite{Barrett2004}. 
Although X-ray attenuation does not directly affect the dark-field signal itself, attenuation can still engender artifacts known as beam-hardening-induced dark-field signals. Since the visibility— and thus the dark-field signal — is dependent on the X-ray beam spectrum, dark-field radiography is similarly susceptible to beam hardening caused by attenuating samples \cite{Pelzer2016, Pandeshwar2020, DeMarco2024}.
In this study, we present a fully automated beam hardening correction framework for clinical dark-field chest radiography, based on deep learning–derived anatomical segmentation. The method leverages rib and clavicle segmentations from a U-Net to generate spatially adaptive attenuation contribution maps. These maps are then used to correct beam hardening-induced dark-field artifacts. We apply and evaluate the approach on a multi-group clinical dataset comprising healthy subjects, as well as subjects with COPD and COVID-19, using both qualitative visual assessment and quantitative metrics to assess artifact suppression and signal homogeneity.

\section{Background}

\subsection{Mathematical Description}
As mentioned in \ref{sec:introduction}, the Talbot-Lau interferometer not only enables the conventional attenuation-based imaging, but also provides access to the dark-field signal, which additionally offers complementary information about the microstructural composition of the irradiated sample. Mathematically, both signal channels can be expressed by \eqref{eq:transmission_monoE} for the transmission $T(E)$ and by \eqref{eq:dark-field_monoE} for the dark-field signal $D(E)$. 

\begin{align}
    &T(E) = \frac{S(E)}{S_0(E)} = \exp \left[-\int^{z_0}_0 \mu(z\mathbf{\hat{e}_\text{z}}, E)\,dz\right] \label{eq:transmission_monoE} \\[1em]
    &D(E) = -\ln\left(\frac{V(E)}{V_0(E)}\right) = c_\text{D}(E)\int^{z_0}_0 \epsilon(z\mathbf{\hat{e}_\text{z}}, E)\,dz \label{eq:dark-field_monoE}
\end{align}

Here, the linear attenuation coefficient $\mu(z\hat{\mathbf{e}}_\text{z},E)$ and the the linear diffusion coefficient $\epsilon(z\hat{\mathbf{e}}_\text{z},E)$ describe the reduction in transmission and visibility, respectively \cite{Bech2010,Strobl2014,Lynch2011,Graetz2020,Yashiro2010}. The unit vector $\hat{\mathbf{e}}_\text{z}$ denotes the direction from the X-ray source to the corresponding detector pixel. The constant $c_\text{D}$ is determined by the grating parameters and the geometric configuration of the setup. The visibility during the sample scan is denoted by $V(E)$, while $V_0(E)$ refers to the visibility in the reference scan without a sample in the beam path. The effective source intensity is given by $S_0(E) = \psi_0(E) \cdot\mathcal{R}(E)$, where $\psi_0(E)$ represents the X-ray intensity and $\mathcal{R}(E)$ is the energy-dependent detector response function.


For a polychromatic X-ray source, the attenuation $A_\text{p}$, visibility $V_\text{p}$, and dark-field signal $D_\text{p}$ are mathematically expressed by \eqref{eq:TM_signal_polyc}, \eqref{eq:visibility_polyc}, and \eqref{eq:DF_signal_polyc} \cite{DeMarco2024}:
\begin{align}
    &A_\text{p} = -\ln\left(\frac{\int_{0}^{\infty}   T(E) \, S_0(E) \, dE}{\int_{0}^{\infty} S_0(E) \, dE}\right)\label{eq:TM_signal_polyc}\\[1em]
    &V_\text{p} = \frac{\int_{0}^{\infty} V(E) \, T(E) \, S_0(E) \,dE}{\int_{0}^{\infty} T(E) \,   S_0(E) \, dE}\label{eq:visibility_polyc}\\[1em]
    &D_\text{p} = -\ln\left(\frac{V_\text{p}}{V_{\text{0,p}}}\right)\label{eq:DF_signal_polyc}.
\end{align}
\subsection*{B. Beam hardening-induced dark-field signal}
\label{subsec:artifacts}
The polychromatic spectrum of medical X-ray sources and the associated energy-dependent interactions with matter cause the X-ray spectrum to be hardened along the beam by the object under examination \eqref{eq:TM_signal_polyc}. This results in a non-linear relationship between measured attenuation and object thickness, a common issue in CT \cite{Barrett2004}.
Although the dark-field signal is based solely on ultra-small-angle scattering (USAXS), which is independent of the attenuation, the spectral distortions induce an indirect effect. Specifically, the beam hardening of the spectrum $S_0(E)$ leads to an artificial dark-field signal -- commonly referred to as beam hardening-induced dark-field signal -- due to the energy dependency of the visibility $V(E)$ and $\epsilon(z\hat{\mathbf{e}}_\text{z},E)$, as described in \eqref{eq:visibility_polyc} \cite{Yashiro2010, Pelzer2016, DeMarco2024}.
Note that the influence of beam-hardening on the dark-field signal also depends on the X-ray spectrum used and the setup parameters and geometry \cite{DeMarco2024}.\\
Since beam hardening of the X-ray spectrum does not convey information about the microstructural composition of the irradiated material, it is desired to remove this artificial signal — particularly pronounced at rib structures and manifesting as a rib-induced step in the dark-field signal — by applying an appropriate beam hardening correction (BHC). \\
If only a single type of attenuator is present, beam hardening can be accurately corrected through calibration measurements \cite{Yashiro2015, bevins_type_2013}. Additionally, it is possible to narrow the source spectrum by applying stronger pre-filtration \cite{Pelzer2016}. However, this approach reduces flux and yields only a modest improvement in beam hardening effects. In chest X-ray imaging, however, these approaches are not applicable, as two different types of attenuators — bone and soft tissue — are present. The key difficulty arises from the fact that the precise composition of bone and soft tissue within the thorax region is unknown. Therefore, a fast and effective method that has shown improvements in image quality in this context is the weighted single-LUT approach (cf. M.E. Lochschmidt \cite{Lochschmidt2025}). In this method, calibration curves were acquired for aluminum and water as surrogates for bone and soft tissue. These were then combined into a single look-up table (LUT) using a weighting factor, which significantly reduced the rib-induced step in the dark-field image. However, in regions of very high attenuation, this approach leads to overcorrection of the dark-field signal. Defining a global weighting factor depends therefor on the specific clinical question or requires a compromise between minimizing the rib artifact and avoiding overcorrection.\\
Building upon the single-LUT approach, the present work takes a significant step forward by implementing a deep learning-based beam hardening correction. This method addresses both the issue of overcorrection and the limitations associated with manually selecting a global weighting factor.

\section{Materials and Methods}
\subsection{Setup}\label{sec:setup}

\begin{figure}[!t]
\centerline{\includegraphics[width=\columnwidth]{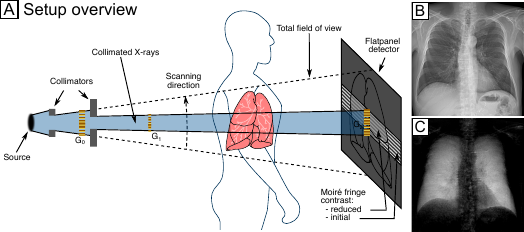}}
\caption{\textbf{Schematic of the clinical dark-field chest radiography system.} The clinical dark-field radiography system integrates a conventional X-ray source, collimators, and a flat panel detector with a Talbot-Lau interferometer comprising a source grating $G_0$, phase grating $G_1$, and analyzer grating $G_2$. This interferometer induces an interference pattern with moiré fringes on the detector plane. By acquiring a series of exposures at varying phase steps of the fringe pattern, both attenuation-based (B) and dark-field (C) radiographs can be simultaneously reconstructed with a perfect temporal and spatial registration. The attenuation-based radiograph is equivalent to a conventional X-ray radiograph, whereas the dark-field image quantifies ultra-small-angle scattering, offering enhanced sensitivity to pulmonary microstructure beyond the resolution of conventional radiography.} 
\label{fig_setup}
\end{figure}

The clinical dark-field prototype, as described in \cite{Willer2021} and \cite{Gassert2021}, such as illustrated in Fig.\ref{fig_setup}, records the attenuation and dark-field images of patients. The system integrates a conventional X-ray source (MRC 200 0508 ROT-GS 1003, Philips Medical Systems) with an aluminum-equivalent pre-filtration of $\SI{2.5}{\milli\meter}$, a collimator box (R 302 MLP/A DHHS, Ralco), three gratings ($\text{G}_0$, $\text{G}_1$, and $\text{G}_2$), and a flat-panel detector (PIXIUM 4343 F4, Trixell). The gratings are mounted on an interferometer arm, which is scanned in the vertical direction, as the gratings do not cover the entire vertical field of view. Moiré fringes are generated in the detector plane due to a slight periodic detuning of $G_2$. Moving the interferometer arm, a stepping curve is then measured for each pixel. A single scan comprises 195 exposures, producing 24 exposures for each pixel, sufficient to generate the stepping curve and so to extract the attenuation and dark-field signals previously described by \eqref{eq:TM_signal_polyc}, and \eqref{eq:DF_signal_polyc}. The total acquisition time for one image is approximately \SI{7}{\second}. All images have been recorded with a tube voltage of $\SI{70}{\kilo\V\text{p}}$, a pulse rate of $\SI{30}{Hz}$, and a pulse duration of $\SI{17.1}{\milli\second}$. For a more detailed description of the image reconstruction methodology, processing pipeline, and correction algorithms, the reader is referred to \cite{Urban2024,Noichl2024}.
The detector features a $\SI{600}{\micro\meter}$ thick CsI scintillator layer and operates with 3$\times$3 pixel binning, resulting in a physical pixel size of $\SI{444}{\micro\meter}$. For study participants positioned at a contact plane $\SI{20}{\centi\meter}$ from the detector, this corresponds to an effective pixel size of approximately $\SI{400}{\micro\meter}$ in the posterior-anterior (p.-a.) orientation. This value may vary slightly depending on patient size and positioning. For a reference patient scanned in the p.-a. position, the effective dose is $\SI{35}{\micro\sievert}$~\cite{Frank2022}. The tube current required to achieve the target detector dose of $\SI{3.75}{\micro\gray}$ at the scanning prototype system is determined for each individual study participant using a calibration curve and the dose information of the conventional system for COPD participants, and a body-mass-index (BMI) correlation curve for COVID-19 participants \cite{Lochschmidt_expo_control_2025}.

\subsection{Selection of Patient Images}
For the methods described below, radiographs taken at the fringe scanning system described in \ref{sec:setup} were used. A total of 51 patient images classified as diseased from the COPD study, 86 from the COVID-19 study, and 37 healthy patients from both studies were selected for the analysis of the BHC method presented here.\\
Spectral CT data were also acquired for two patients.

\subsection{Segmentation}
\begin{figure*}[!t]
\centering
\includegraphics[width=\textwidth]{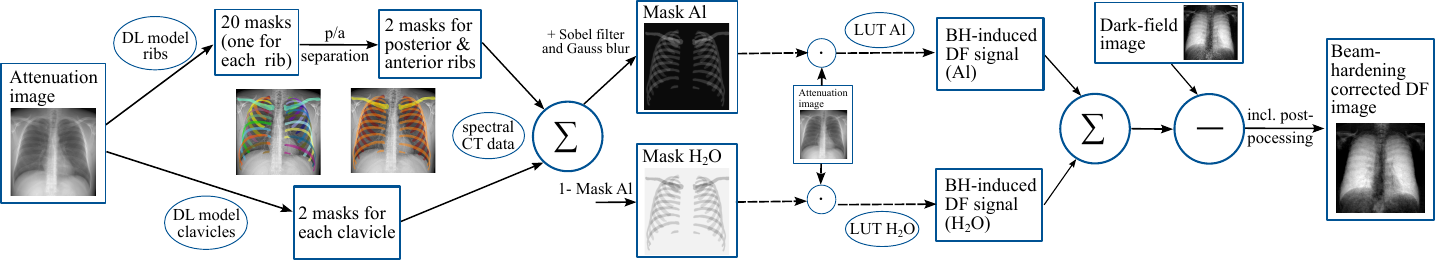}
\caption{\textbf{Schematic overview of the proposed beam-hardening correction pipeline.} Deep learning-based models are used to segment ribs and clavicles from the attenuation image, followed by separation into anterior and posterior components. Material-specific masks are generated using spectral CT-based decomposition into aluminum and water. Beam-hardening-induced dark-field signals are estimated via calibrated look-up tables (LUTs) and refined through Gaussian smoothing and Sobel filtering. The resulting artifact maps are subtracted from the raw dark-field image.  
}
\label{fig:overview}
\end{figure*}

\subsubsection{Rib Segmentation}

Automated rib segmentation was performed using a deep convolutional neural network based on a U-Net architecture with skip connections and an EfficientNet-B0 encoder pre-trained on ImageNet. The model was trained on the publicly available VinDr-RibCXR dataset~\cite{Nguyen2021}, comprising 245 chest radiographs with manually annotated masks for the first ten ribs bilaterally. The dataset was split into 196 training and 49 validation images.\\
A total of 16 attenuation radiographs acquired at the dark-field setup were selected and manually segmented with an internal annotation tool to test the performance of the model.. In each image, the first ten ribs on both sides were labeled individually, resulting in 20 rib masks per image, consistent with the structure of the training data.\\
The network was implemented in PyTorch (v2.0.1) and trained using the Adam optimizer and Dice loss function on an NVIDIA RTX A4000 GPU. Training parameters are summarized in Table~\ref{tab:param}.
\begin{table}[!h]
    \centering
    \caption{Training parameters for rib and clavicle segmentation}
    \begin{tabular}{c||c c c c}
        Parameters & Learning rate & Epochs & Batch size & \#Classes\\
        \hline
        Rib & $1e-3$ & 195 & 16 & 20\\
        Clavicle & $5e-4$ & 274 & 8 & 2
    \end{tabular}
    \label{tab:param}
\end{table}

\subsubsection{Clavicle Segmentation}
Clavicle segmentation was conducted using a dataset of 80 attenuation images acquired with the clinical dark-field scanner.
Manual annotations were performed using the MATLAB segmentation toolbox (MathWorks, v9.14, R2023a). The dataset was stratified into 56 training, 12 validation, and 12 testing images.\\
The segmentation model employed the same U-Net architecture and encoder as used for rib segmentation. The model was trained in PyTorch using the Adam optimizer and Dice loss. Training parameters were optimized for this task and are summarized in Table~\ref{tab:param}.\\
This segmentation model enables automated segmentation of the clavicles in attenuation images acquired from the dark-field prototype, thereby supporting anatomically informed beam hardening correction.


\subsection{Attenuation Contribution Mask Creation}
\subsubsection{Material Decomposition}\label{sec:Material_decomposition}
To derive material-specific attenuation contributions, spectral raw data were acquired using a dual-energy CT (IQon spectral CT, Philips Healthcare, Best, The Netherlands). Virtual Mono-energetic images (VMIs) at $\SI{50}{\kilo\eV}$ and $\SI{200}{\kilo\eV}$ were then generated for one male and one female patient using IntelliSpace Portal (version 12.1, Philips Healthcare). Using both VMIs of each patient, a material decomposition into water- and aluminum-attenuation images was performed for each, serving as surrogates for soft tissue and bone, respectively. The water and aluminum attenuation maps were normalized to express relative contributions to total attenuation.\\
Cross-sectional profiles through the ribs and clavicles in the decomposed attenuation maps were analyzed to derive region-specific contribution values, differences between anterior and posterior ribs, and differences between male and female patients.

\subsubsection{Attenuation Contribution Mask Generation}
Anatomically resolved attenuation masks were generated by applying the attenuation contribution weights to the segmentation masks.
Posterior and anterior rib components were separated by identifying the overlap region at the lung boundary; pixels superior to this boundary were assigned to posterior ribs, whereas those inferior were assigned to anterior ribs.
To incorporate the increased attenuation at bone edges observed in spectral CT, a Sobel filter was applied to the binary masks to extract rib edges, which were subsequently dilated and incorporated into the attenuation map with a weighting factor of 0.05. 
The aluminum attenuation contribution map was smoothed using a Gaussian kernel ($\sigma=2.2$, truncated at $3\sigma$) to ensure gradual spatial transitions and prevent sharp discontinuities in the correction process. The complementary water mask was obtained as
\begin{align}
    \omega_\mathrm{H_2O}(x, y) = 1 - \omega_\mathrm{Al}(x, y).\label{eq:water_weighting}
\end{align}

The two most caudal ribs, not represented in the training data, were excluded from the segmentation. As these ribs primarily overlay abdominal structures, their omission was not expected to affect lung-specific beam hardening correction.

\subsubsection{Beam Hardening Correction Processing}
\label{sec:bhc}

The beam hardening correction (BHC) procedure builds upon the single LUT method introduced by Lochschmidt et al~\cite{Lochschmidt2025}, where the beam hardening-induced dark-field signal is corrected using LUTs for water and aluminum and averaged with a global aluminum weighting factor and a correction-related bias correction term. Both have to be chosen depending on the clinical question. Since the bias correction term is theoretically dependent on how large the attenuation of pure water is in the image, but the exact position of these areas is not known, e.g. in the form of a mask, a global value must be defined that is either a compromise of the entire attenuation range or only optimizes the correction for regions of equal attenuation. An optimization for the upper lung region would therefore lead to overcorrection in the heart region, and one needs to choose new parameters to optimize for the heart region, which would mean an undercorrection in the upper lung regions.\\
In the approach proposed here, this global weighting factor and the bias-correction term are replaced with a spatially resolved weighting derived from attenuation contribution masks described above in combination with the deep learning-based bone segmentation. Specifically, the aluminum and water masks are used to compute a pixel-wise weighted sum of the LUT-based beam hardening-induced dark-field signals. The resulting correction map is then subtracted from the raw dark-field image to obtain the corrected image. Fig.~\ref{fig:overview} summarizes the complete pipeline, including segmentation of ribs and clavicles, p.-a. separation, mask generation using spectral CT-based attenuation contributions, and the final BHC implementation using pixel-wise LUT interpolation and subtraction.

\subsection{Statistical Analysis}\label{subsect:statistical_analysis}
All analyses were performed using Python (v3.8.10; NumPy v1.24.4, SciPy v1.10.1, pandas
v2.0.3) and R (v4.5.0). To quantify the impact of the BHC, the sum of the dark-field extinction coefficient was calculated for each patient's lung mask and finally normalized by the lung volume. The lung volume was calculated by a model using lateral and p.-a. orientations of each patient \cite{Pierce1979}.\\
The effectiveness of beam hardening correction (BHC) in improving the homogeneity of the lung dark-field signal was evaluated using the coefficient of variation (CV) \eqref{eq:metrics_CV}, the interquartile range (IQR) \eqref{eq:metrics_IQR} \cite{box2005statistics}, and the overlap of two IQRs \eqref{eq:overlap} computed within manually defined lung masks. 
\begin{align}
    &\text{CV} = \frac{\sigma^2}{\mu}\label{eq:metrics_CV}\\[1em]
    &\text{IQR} = Q_3 - Q_1 \label{eq:metrics_IQR}\\[1em]
    &\delta_{\text{IQR(B)}}^{\text{IQR(A)}} = \max\Big(0,\ 
    \min\left(Q_{3}^{(A)}, Q_{3}^{(B)}\right) \notag \\
    & \hspace{6.5em}- \max\left(Q_{1}^{(A)}, Q_{1}^{(B)}\right)\Big) \label{eq:overlap}
\end{align}
where $\sigma^2$ is the variance and $\mu$ is the mean value. $Q_1$ separates the bottom 25\% of the data from the rest and $Q_3$ the top 25\% from the rest.

For each subject, CV and IQR were computed across the lung region in p.-a. orientation, both before and after application of the segmentation-based BHC. Lower values in these metrics correspond to increased signal homogeneity and thus indicate reduced beam hardening artifacts.\\
To assess the statistical significance of the CV reduction after BHC, the Wilcoxon signed-rank test  was applied to paired CV values (no BHC vs. BHC) within each clinical cohort (COPD, COVID-19, Healthy). This non-parametric test does not assume normality and was used with a two-sided significance threshold of $p < 0.05$.\\
Group-wise CV distributions were visualized using box plots. Percentage reductions were summarized using both mean and median values. The IQR was used descriptively to quantify signal spread.

\section{Results}
\subsection{Attenuation Contribution Mask}
As described in \ref{sec:Material_decomposition}, cross-sectional profiles through the ribs and clavicles in the decomposed maps were performed and resulted in characteristic values for the posterior ribs, anterior ribs, clavicle, and the edges of the ribs. Furthermore, a uniform background contribution of 0.05 was measured in regions between ribs and incorporated into the weighting model. Table~\ref{tab:al_contribution} summarizes the empirically derived aluminum weights.
\begin{table}[!h]
    \centering
    \setlength{\tabcolsep}{5pt}
    \caption{Aluminum weights for several bone regions and the background.}
    \begin{tabular}{c||c c c c c}
        Aluminum & Posterior & Anterior & Clavicle & Edge & Background\\
        \hline
        Male & 0.15 & 0.1 & 0.2 & 0.05 & 0.05\\
        Female & 0.1 & 0.05 & 0.2 & 0.05 & 0.05
    \end{tabular}
    \label{tab:al_contribution}
\end{table}

\subsection{Qualitative Evaluation}
\begin{figure*}[!t]\centerline{\includegraphics[width=\textwidth]{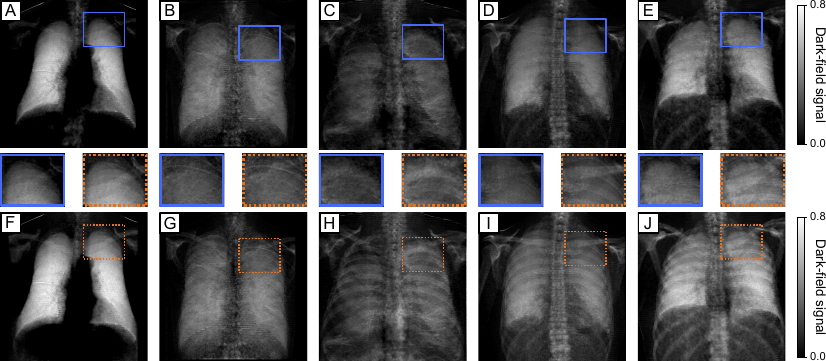}}
\caption{\textbf{Effect of segmentation-based beam hardening correction (BHC) across representative clinical cases.}  
Dark-field chest radiographs of five representative patients are shown with (A--E) and without (F--J) the proposed segmentation-based BHC. (A, F) Healthy subjects (B, G) and (C, H) patients with chronic obstructive pulmonary disease (COPD); (D, I) and (E, J) patients with COVID-19 pneumonia. The segmentation-based BHC substantially reduces rib- and clavicle-induced artifacts, yielding a more homogeneous dark-field signal across the lung fields. Insets highlight the regions corresponding to the segmented ribs and clavicles, illustrating the suppression of beam hardening-induced dark-field in osseous structures.}
\label{fig:bhc_comparison}
\end{figure*}

The proposed segmentation-based BHC described was applied to 174 clinical dark-field chest radiographs from patients not included in the segmentation model's training or validation cohorts. Representative results from five patients -- one healthy subject, two with COPD of varying severity, and two with COVID-19 pneumonia -- are shown in Fig. \ref{fig:bhc_comparison}.\\
In the uncorrected images, pronounced beam hardening-induced signals from osseous structures, particularly ribs and clavicle, are visible across the lung. These structured artifacts introduce marked inhomogeneity and obscure the overlapping lung parenchyma. Following BHC, these bone-induced features are substantially reduced, yielding a more homogeneous signal distribution and improving visualization of the pulmonary microstructure. Insets highlight the segmented rib and clavicle regions, further illustrating artifact suppression by the BHC.

\subsection{Quantitative Evaluation}
\begin{figure}[!h]
\centerline{\includegraphics[width=\columnwidth]{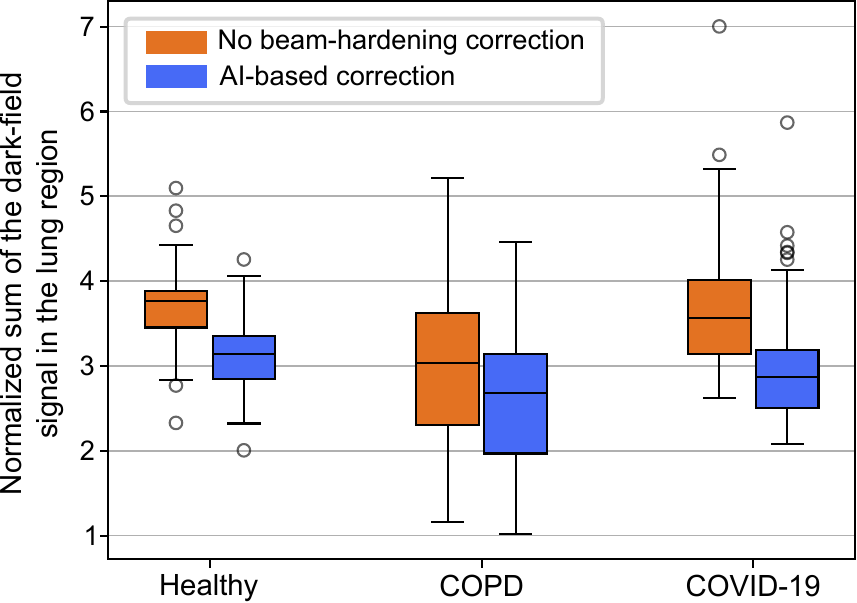}}
\caption{\textbf{Quantitative comparison of beam hardening correction (BHC) approaches in clinical dark-field chest radiography.} The plot shows the sum of the dark-field signal in the lung region, normalized by lung volume, for three patient cohorts: Healthy, chronic obstructive pulmonary disease (COPD), and COVID-19. Results are shown without a correction of beam-hardening effects (orange) and with the proposed AI-Segmentation-Based BHC (blue). The segmentation-based approach results in consistently higher normalized dark-field signals across all cohorts, indicating improved recovery of true lung signal and reduction of beam hardening artifacts from bones. Each box indicates the interquartile range with the median, and whiskers extend to 1.5× the IQR; outliers are shown as circles. The statistical evaluation described in \ref{subsect:statistical_analysis} is shown in Tab.\ref{tab:statistical_evaluation_results}.}
\label{fig:boxplots}
\end{figure}

Group-wise comparisons were performed across healthy, COPD, and COVID-19 cohorts by assessing the total volume obtained from the normalized sum of the dark-field signal described in \ref{subsect:statistical_analysis}.
Figure~\ref{fig:boxplots} shows the resulting distributions for the uncorrected signal and the proposed segmentation-based BHC. 

Across all cohorts,  segmentation-based BHC resulted in a lower normalized dark-field signal. In healthy subjects, the mean signal decreased from 3.76 to 3.14 ($-16.5\%$), in COPD patients from 3.04 to 2.68 ($-11.8\%$), and in COVID-19 patients from 3.57 to 2.87 ($-19.6\%$).\\
In addition to reductions of the normalized dark-field signal, diagnostic class separability improved. The median overlap between healthy and COPD patients decreased from 2.77 to 2.25 ($-18.8\%$), and between healthy and COVID-19 patients from 2.48 to 2.17 ($-12.5\%$). Both reductions were statistically significant ($p < 0.01$, Wilcoxon signed-rank test).

\begin{table*}[!h]
    \centering
    \setlength{\tabcolsep}{5pt}
    \caption{Statistical analysis of the AI segmentation-based method described in \ref{subsect:statistical_analysis}. The results or data this analysis refers to are also visualized in Fig.\ref {fig:boxplots} as boxplots. Here, $\Tilde{x}$ represents the median value. The coefficient of variation (CV) and the overlap of two interquartile ranges of two different boxplots $\delta$ were calculated with \eqref{eq:metrics_CV} and \eqref{eq:overlap}, respectively.}
    \begin{tabular}{l||ccc|cc|ccc}
        Parameters & $\Tilde{x}_{\text{Healthy}}$ & $\Tilde{x}_{\text{COPD}}$ & $\Tilde{x}_{\text{COVID-19}}$& $\delta_{\text{IQR(COPD)}}^{\text{IQR(Healthy)}}$ & $\delta_{\text{IQR(COVID-19)}}^{\text{IQR(Healthy)}}$& $\text{CV}_{\text{Healthy}}$& $\text{CV}_{\text{COPD}}$& $\text{CV}_{\text{COVID-19}}$\\
        \hline
        Values (no BHC) &3.76  &3.04  &3.57&2.77 &2.48& 0.32&0.31&0.38\\
        \hline
        Values (AI) &3.14  &2.68 &2.87& 2.25&2.17&0.27&0.26&0.31\\
        \hline
        \hline
        Relative Change & $\textbf{-16.5\%}$ & $\textbf{-11.8\%}$&$\textbf{-19.6\%}$&$\textbf{-18.8\%}$ &$\textbf{-12.5\%}$&$\textbf{-15.6\%}$&$\textbf{-16.1\%}$&$\textbf{-18.4\%}$ \\
        Wilcoxon test (p-value) & - & - & - & $<0.01$ & $<0.01$ & $< 0.01$ & $<0.01$ & $<0.01$
    \end{tabular}
    \label{tab:statistical_evaluation_results}
\end{table*}

To further evaluate intra-patient signal homogeneity, the CV was computed individually for each patient based on their lung segmentation. 
As shown in Table \ref{tab:statistical_evaluation_results}, segmentation-based BHC leads to a consistent and significant CV reduction across all three cohorts. In healthy subjects, CV decreased from 0.32 to 0.27; in COPD patients, from 0.31 to 0.26; and in COVID-19 patients, from 0.38 to 0.31. All changes were statistically significant ($p < 0.001$, Wilcoxon signed-rank test), indicating improved spatial uniformity of the lung dark-field signal.

\section{Discussion}
In grating-based dark-field radiography, highly attenuating anatomical regions such as ribs and clavicles introduce artifacts due to the beam hardening induced dark-field signal. These artifacts obscure lung parenchyma and reduce both the interpretability and quantitative reliability of dark-field chest radiographs. \\ 
To address this limitation, a segmentation-based beam hardening correction (BHC) framework was developed and evaluated in a clinical setting. The method combines deep learning-based anatomical segmentation of osseous structures and material-specific attenuation weighting derived from dual-energy spectral CT decomposition with LUT-based methods published before \cite{Lochschmidt2025}. This enables regionally adaptive correction of beam hardening-induced dark-field signal and prevents overcorrections within the entire dark-field images while preserving true microstructural signals from lung tissue.\\ 
Applied to a clinical cohort of 174 patients, the proposed method demonstrated consistently improved image quality and quantitative characteristics across healthy subjects and patients with chronic COPD and COVID-19 pneumonia.\\
The correction significantly reduced rib- and clavicle-induced artifacts, yielding more homogeneous dark-field representations of the lungs and improving the visibility of parenchymal regions. Quantitative analysis confirmed a statistically significant reduction in volume-normalized dark-field signal variation across all cohorts. More importantly, diagnostic separability improved: distributional overlap between healthy and COPD patients decreased by 18.8\%, and between healthy and COVID-19 patients by 12.5\% (both $p < 0.01$, Wilcoxon signed-rank test). 
These findings support improved classification potential for disease detection and differentiation. \\ 
Beyond inter-group discrimination, the method also improved intra-patient signal consistency markedly. The CV within individual lung masks decreased significantly across all cohorts ($p < 0.01$), indicating enhanced spatial homogeneity of the corrected dark-field signal. \\
The correction also increased the contrast between healthy lung parenchyma and diseased lungs, potentially improving sensitivity in early disease detection.  \\
Several limitations should be acknowledged. The performance of the proposed framework is dependent on the accuracy of the segmentation model and rib separation algorithm. Segmentation errors, especially in patients with atypical anatomy or severe pathology, may lead to local over- or under-correction. Furthermore, attenuation contribution weights were derived from only two spectral CT datasets, limiting generalizability. Metallic implants or foreign bodies were not addressed, as they are not currently included in the segmentation or decomposition models.\\ 
Future work should explore the integration of spectral imaging into the clinical dark-field system. Dual-layer, photon-counting detectors or rapid kVp-switching systems can provide direct, pixel-wise material decomposition without reliance on segmentation. These detectors or acquisition types enable the acquisition of spectrally resolved X-ray data, allowing basis-material decomposition into soft tissue and bone or water and aluminum, respectively. Such an approach would enable fully patient-specific correction and eliminate dependence on anatomical priors or training data. Prior work has demonstrated the feasibility of spectral models for dark-field signal characterization \cite{Taphorn2023,Sellerer2021}, including differentiation of diseases such as emphysema and fibrosis \cite{Taphorn2021}. Integrating these models into the correction pipeline could further improve robustness, generalizability, and clinical applicability.

\section{Conclusion}
This study presents the first fully automated, anatomically adaptive beam hardening correction framework for clinical X-ray dark-field chest radiography. The proposed method combines deep learning–based segmentation of ribs and clavicles with material-specific attenuation weighting to suppress beam hardening artifacts introduced by osseous structures. The correction improves intra-pulmonary signal homogeneity, enhances diagnostic separability between healthy and diseased lungs, and supports both qualitative and quantitative interpretation of dark-field images. These improvements address a major limitation of dark-field imaging and support its further integration into clinical research and diagnostic workflows.

\appendices

\printbibliography

\end{document}